\documentclass[ALICE,manyauthors]{cernphprep}

\usepackage[comma,square,numbers,sort&compress]{natbib}
\usepackage{hyperref}
\usepackage{lineno}
\usepackage{xspace}
\usepackage{mathrsfs}
\usepackage[T1]{fontenc}
\begin{document}
%

\newcommand{\pp}           {pp\xspace}
\newcommand{\ppbar}        {\mbox{$\mathrm {p\overline{p}}$}\xspace}
\newcommand{\XeXe}         {\mbox{Xe--Xe}\xspace}
\newcommand{\PbPb}         {\mbox{Pb--Pb}\xspace}
\newcommand{\pA}           {\mbox{pA}\xspace}
\newcommand{\pPb}          {\mbox{p--Pb}\xspace}
\newcommand{\AuAu}         {\mbox{Au--Au}\xspace}
\newcommand{\dAu}          {\mbox{d--Au}\xspace}

\newcommand{\s}            {\ensuremath{\sqrt{s}}\xspace}
\newcommand{\snn}          {\ensuremath{\sqrt{s_{\mathrm{NN}}}}\xspace}
\newcommand{\pt}           {\ensuremath{p_{\rm T}}\xspace}
\newcommand{\meanpt}       {$\langle p_{\mathrm{T}}\rangle$\xspace}
\newcommand{\ycms}         {\ensuremath{y_{\rm CMS}}\xspace}
\newcommand{\ylab}         {\ensuremath{y_{\rm lab}}\xspace}
\newcommand{\etarange}[1]  {\mbox{$\left | \eta \right |~<~#1$}}
\newcommand{\yrange}[1]    {\mbox{$\left | y \right |~<~#1$}}
\newcommand{\dndy}         {\ensuremath{\mathrm{d}N_\mathrm{ch}/\mathrm{d}y}\xspace}
\newcommand{\dndeta}       {\ensuremath{\mathrm{d}N_\mathrm{ch}/\mathrm{d}\eta}\xspace}
\newcommand{\avdndeta}     {\ensuremath{\langle\dndeta\rangle}\xspace}
\newcommand{\dNdy}         {\ensuremath{\mathrm{d}N_\mathrm{ch}/\mathrm{d}y}\xspace}
\newcommand{\Npart}        {\ensuremath{N_\mathrm{part}}\xspace}
\newcommand{\Ncoll}        {\ensuremath{N_\mathrm{coll}}\xspace}
\newcommand{\dEdx}         {\ensuremath{\textrm{d}E/\textrm{d}x}\xspace}
\newcommand{\RpPb}         {\ensuremath{R_{\rm pPb}}\xspace}
\newcommand{\pT}           {\ensuremath{p_{\rm T,e}}\xspace}
\newcommand{\pTee}         {\ensuremath{p_{\rm T,ee}}\xspace}
\newcommand{\mee}         {\ensuremath{m_{\rm ee}}\xspace}

\newcommand{\nineH}        {$\sqrt{s}~=~0.9$~Te\kern-.1emV\xspace}
\newcommand{\seven}        {$\sqrt{s}~=~7$~Te\kern-.1emV\xspace}
\newcommand{\twoH}         {$\sqrt{s}~=~0.2$~Te\kern-.1emV\xspace}
\newcommand{\twosevensix}  {$\sqrt{s}~=~2.76$~Te\kern-.1emV\xspace}
\newcommand{\five}         {$\sqrt{s}~=~5.02$~Te\kern-.1emV\xspace}
\newcommand{\twosevensixnn}{$\sqrt{s_{\mathrm{NN}}}~=~2.76$~Te\kern-.1emV\xspace}
\newcommand{\fivenn}       {$\sqrt{s_{\mathrm{NN}}}~=~5.02$~Te\kern-.1emV\xspace}
\newcommand{\LT}           {L{\'e}vy-Tsallis\xspace}
\newcommand{\GeVc}         {Ge\kern-.1emV/$c$\xspace}
\newcommand{\MeVc}         {Me\kern-.1emV/$c$\xspace}
\newcommand{\TeV}          {Te\kern-.1emV\xspace}
\newcommand{\GeV}          {Ge\kern-.1emV\xspace}
\newcommand{\MeV}          {Me\kern-.1emV\xspace}
\newcommand{\GeVmass}      {Ge\kern-.2emV/$c^2$\xspace}
\newcommand{\MeVmass}      {Me\kern-.2emV/$c^2$\xspace}
\newcommand{\lumi}         {\ensuremath{\mathcal{L}}\xspace}

\newcommand{\ITS}          {\rm{ITS}\xspace}
\newcommand{\TOF}          {\rm{TOF}\xspace}
\newcommand{\ZDC}          {\rm{ZDC}\xspace}
\newcommand{\ZDCs}         {\rm{ZDCs}\xspace}
\newcommand{\ZNA}          {\rm{ZNA}\xspace}
\newcommand{\ZNC}          {\rm{ZNC}\xspace}
\newcommand{\SPD}          {\rm{SPD}\xspace}
\newcommand{\SDD}          {\rm{SDD}\xspace}
\newcommand{\SSD}          {\rm{SSD}\xspace}
\newcommand{\TPC}          {\rm{TPC}\xspace}
\newcommand{\TRD}          {\rm{TRD}\xspace}
\newcommand{\VZERO}        {\rm{V0}\xspace}
\newcommand{\VZEROA}       {\rm{V0A}\xspace}
\newcommand{\VZEROC}       {\rm{V0C}\xspace}
\newcommand{\Vdecay} 	   {\ensuremath{V^{0}}\xspace}

\newcommand{\ee}           {\ensuremath{e^{+}e^{-}}} 
\newcommand{\pip}          {\ensuremath{\pi^{+}}\xspace}
\newcommand{\pim}          {\ensuremath{\pi^{-}}\xspace}
\newcommand{\kap}          {\ensuremath{\rm{K}^{+}}\xspace}
\newcommand{\kam}          {\ensuremath{\rm{K}^{-}}\xspace}
\newcommand{\pbar}         {\ensuremath{\rm\overline{p}}\xspace}
\newcommand{\kzero}        {\ensuremath{{\rm K}^{0}_{\rm{S}}}\xspace}
\newcommand{\lmb}          {\ensuremath{\Lambda}\xspace}
\newcommand{\almb}         {\ensuremath{\overline{\Lambda}}\xspace}
\newcommand{\Om}           {\ensuremath{\Omega^-}\xspace}
\newcommand{\Mo}           {\ensuremath{\overline{\Omega}^+}\xspace}
\newcommand{\X}            {\ensuremath{\Xi^-}\xspace}
\newcommand{\Ix}           {\ensuremath{\overline{\Xi}^+}\xspace}
\newcommand{\Xis}          {\ensuremath{\Xi^{\pm}}\xspace}
\newcommand{\Oms}          {\ensuremath{\Omega^{\pm}}\xspace}
\newcommand{\degree}       {\ensuremath{^{\rm o}}\xspace}

\begin{titlepage}
\PHyear{2020}       
\PHnumber{095}      
\PHdate{28 May}  


\title{Soft-dielectron excess in proton--proton collisions at $\mathbf{\sqrt{\textit s}=13}$~TeV}
\ShortTitle{Soft-dielectron excess in proton--proton collisions at $\sqrt{\textit s}=13$~TeV}   

\Collaboration{ALICE Collaboration\thanks{See Appendix~\ref{app:collab} for the list of collaboration members}}
\ShortAuthor{ALICE Collaboration} 

\begin{abstract}

A measurement of dielectron production in proton--proton (pp) collisions at $\sqrt{s} = 13$ TeV, recorded with the \mbox{ALICE} detector at the CERN LHC, is presented in this Letter. The data set was recorded with a reduced magnetic solenoid field. This enables the investigation of a kinematic domain at low dielectron invariant mass \mee and pair transverse momentum \pTee that was previously inaccessible at the LHC. The cross section for dielectron production is studied as a function of  \mee, \pTee, and event multiplicity ${\rm d} N_{\rm ch}/{\rm d} \eta$. The expected dielectron rate from hadron decays, called hadronic cocktail, utilizes a parametrization of the measured $\eta/\pi^0$ ratio in pp and proton-nucleus (p--A) collisions, assuming that this ratio shows no strong dependence on collision energy at low transverse momentum. Comparison of the measured dielectron yield to the hadronic cocktail at $0.15<\mee<0.6$~GeV/$c^2$ and for $\pTee < 0.4$ GeV/$c$ indicates an enhancement of soft dielectrons, reminiscent of the 'anomalous' soft-photon and -dilepton excess in hadron--hadron collisions reported by several experiments under different experimental conditions. The enhancement factor over the hadronic cocktail amounts to $1.61\pm 0.13\,(\rm{stat.})\pm 0.17\,(\rm{syst., data}) \pm 0.34\,(\rm{syst., cocktail})$ in the \mbox{ALICE} acceptance. Acceptance-corrected excess spectra in \mee and \pTee are extracted and compared with calculations of dielectron production from hadronic bremsstrahlung and thermal radiation within a hadronic many-body approach.

\end{abstract}
\end{titlepage}

\setcounter{page}{2} 


The study of lepton pair production is an important tool to investigate the properties of hadronic and nuclear collisions
as they can leave the strongly interacting system at any stage of its evolution. 
In order to single out possible medium contributions to the dilepton yield in nucleus--nucleus collisions on top of those from hadron decays, 
studies in hadronic collision systems are instrumental to obtain a medium-free reference.
Recent measurements of dielectron (e$^+$e$^-$) production at midrapidity in proton--proton (pp) collisions at the Large Hadron Collider (LHC) at CERN~\cite{ALICE_dielecton7TeV,ALICE_dielecton13TeV,ALICE_dielectron5TeV} and at the Relativistic Heavy-Ion Collider (RHIC) at BNL~\cite{PHENIX_pp, PHENIX_pp2,STAR_pp} are compatible with the expectations from hadron decays, i.e.,\,with the 
hadronic cocktail, and show no indication of medium effects within the experimental uncertainties.
In contrast to this, recent measurements 
of hadronic observables in small collision systems at the LHC~\cite{pp_collectivity1,ALICE_BlastWave,pp_collectivity3, ATLAS} and at RHIC~\cite{Phenix_collectivity,Phenix_corrSys,Phenix_corrEnergy} reveal signs of collectivity and equilibration of the final-state particles at high multiplicities. This suggests that considerable interaction in an intermediate state may indeed be at work even in pp collisions, which should also give rise to the emission of electromagnetic radiation.

The production of soft photons in hadronic collision systems was extensively studied in fixed-target experiments at beam momenta ranging from 10.5 to 450 GeV/$c$. 
Except for the lowest collision energies~\cite{SoftPhoton4}, most experiments reported an excess of soft photons compared with the expectation from hadron decays that could not be explained by initial- and final-state bremsstrahlung~\cite{SoftPhoton1,SoftPhoton2,SoftPhoton3}. The emergence of a photon excess in a transverse momentum (\pt) range far below 0.2 GeV/$c$ was dubbed the {\em soft-photon puzzle} because bremsstrahlung from initial- and final-state particles should dominate over the radiation from any intermediate state in the soft limit, as stated by the Low theorem~\cite{LowTheorem}.  
This raised speculations about the existence of a radiating intermediate state with characteristic time and length scales well above 1~fm~\cite{Schukraft}; a scenario that can be largely ruled out by more recent measurements of the source size in pp collisions from particle interferometry~\cite{ALICE_femto,ALICE_femto2,STAR_femto}. Several possible mechanisms were proposed to explain the observations, including the annihilation of soft partons~\cite{SoftAnn1,SoftAnn2,SoftAnn3,SoftAnn4,SoftAnn5,SoftAnn6}, the production of a cold non-equilibrium state of quarks and gluons~\cite{VanHove_ColdQGP,VanHove_ColdQGP2}, and the emission of synchrotron radiation off quarks that are accelerated in the chromomagnetic fields of the colliding hadrons~\cite{Nachtmann_QCDsynchrotron,Botz_QCDsynchrotron}. A final conclusion on the interpretation of the soft-photon excess has not been reached though~\cite{SoftPhotonReview,SoftPhotonReview2}.

In the dilepton sector, an enhancement over the hadronic cocktail was observed for 
both electron and muon pairs at small invariant masses in pp collisions at the Intersecting Storage Rings (ISR)~\cite{AFS_Vincent}, and in fixed-target experiments with $\pi$ and p beams from 10 to 400 GeV/$c$~\cite{SoftDielectron2,SoftDielectron3,SoftDielectron4,SoftDielectron5,SoftDielectron6,SoftDimuon1,SoftDimuon2,SoftDimuon3,SoftDimuon4,SoftDimuon5,SoftDimuon6}. Similarly to the case of real photons, the excess yield could not be reconciled with the expectation from hadronic bremsstrahlung. These observations are supported by findings of an enhanced e$^+/\pi$ ratio at the ISR~\cite{AFS_EoverPi}. However, the observations in the dilepton sector remained controversial because other experiments reported results that were compatible with bremsstrahlung and hadron decays only~\cite{CERESbaseline2,SoftDielectron1,HELIOSDileptons}. The question of anomalous soft-dilepton production in hadronic collisions awaits further experimental input since three decades.

In a dedicated campaign during pp operation at $\sqrt{s}=13$~TeV, the \mbox{ALICE} central-barrel detectors~\cite{ALICE_detector3} were operated inside a lower magnetic solenoid field, which increased the sensitivity for electrons at low \pt (the term 'electron' is used here for electrons and positrons). This makes a reassessment of soft dielectron production possible that could not be performed in a previous analysis at nominal field~\cite{ALICE_dielecton13TeV}. 

A detailed description of the \mbox{ALICE} apparatus and its performance can be found in~\cite{ALICE_detector1}. 
The tracking of charged particles is performed by the Inner Tracking System (ITS)~\cite{ITS} and by the Time Projection Chamber (TPC)~\cite{TPC}, which are located in the central barrel and are surrounded by a solenoid, providing a homogeneous magnetic field along the beam direction. 
The TPC is used for particle identification (PID) via the measurement of the specific ionization energy loss (d$E$/d$x$). Additional PID information is provided by the Time-Of-Flight (TOF)~\cite{TOF} system.
Collision events are selected using the V0 detectors located on either side of the interaction point. Furthermore, the events are classified on the basis of the V0 signal amplitude. The event classes are reported in terms of ${\rm d} N_{\rm ch}/{\rm d} \eta$ at midrapidity~\cite{ALICE_PionMult}.

The data samples analyzed for this Letter were recorded in 2016--2018 in pp collisions at $\sqrt{s}=13$~TeV with \mbox{ALICE}, employing a setup where the magnetic solenoid field was reduced from 0.5~T to 0.2~T. This increases the acceptance and efficiency of the tracking and TOF detectors, extending the single electron selection from \mbox{$\pT \geq 0.2$~GeV/$c$} down to \mbox{$\pT \geq 0.075$~GeV/$c$} and 
providing access down to pair transverse momenta \mbox{$\pTee \geq 0$} for invariant masses $\mee>0.15$~GeV/$c^2$.
The minimum bias (MB) event trigger is constructed using a coincident signal in both V0 scintillators.
Interaction vertices are reconstructed by extrapolation of ITS track segments towards the nominal interaction point. Events with multiple reconstructed vertices are tagged as pile-up and rejected. The requirement on the vertex position to be within $\pm10$ cm of the nominal interaction point in 
beam direction is employed to ensure a uniform detector performance. After event selection, a total of $5.42\times10^8$ MB pp events remain for further analysis, corresponding to an integrated luminosity of $\mathscr{L_{\mathrm{int}}} = 9.38\pm 0.47$~nb$^{-1}$
based on the visible cross section observed by the V0 trigger extracted from a van der Meer scan~\cite{ALICE_VanDerMeer}.

The electron candidates used in this analysis are selected in the transverse momentum range $\pT>0.075$ GeV/$c$ and pseudorapidity $|\eta_{\rm e}| < 0.8$.
Further track and PID selection criteria are 
identical to those described in~\cite{ALICE_dielecton13TeV} with the exception of a stronger requirement on the maximum distance of closest approach (DCA) to the primary vertex in the longitudinal direction (DCA$_\mathrm{z}< 0.3$  cm) to remove a contribution of looping tracks in the TPC.

Since pairs of electrons originating from the same source cannot be identified unambiguously, a statistical approach is applied to extract the yield of correlated pairs. 
To this end, a combinatorial pairing of all electron candidates in an event is performed.
Additional photon conversion rejection is achieved by removing pairs based on their characteristic orientation relative to the magnetic field~\cite{ALICE_dielecton7TeV}.

The combinatorial background estimate is constructed from same-event pairs with the same charge sign, corrected for charge-dependent acceptance effects, and subtracted from the opposite-sign pair distribution, following the approach described in~\cite{ALICE_dielecton13TeV}. 
To correct the signal for the finite reconstruction efficiency, a Monte Carlo (MC) simulation is used as described in~\cite{ALICE_dielecton13TeV}.
Proton--proton events are generated using the Monash 2013 tune of PYTHIA 8.1~\cite{PYTHIA8_Monash} to simulate light-hadron decays,
while the Perugia 2011 tune of PYTHIA 6.4~\cite{PYTHIA6} is utilized to embed heavy-flavor hadrons that decay to electrons.
The generated particles are propagated through the detector using GEANT~3~\cite{GEANT3}. The final efficiency as a function of \mee and \pTee is the average of the efficiencies of the different dielectron sources, weighted by their expected contribution to the hadronic cocktail (see below).     

The systematic uncertainties of the data are evaluated as described in~\cite{ALICE_dielecton13TeV} by simultaneous variation of the single-electron tracking and PID selection criteria. The track sample is varied by changing the criteria on the number of space points in TPC and ITS, the $\chi^2$ of the track fits, and the criteria used for electron selection and hadron rejection. These variations imply changes of the pair efficiency by up to about 30\%.
The systematic uncertainty is calculated as the root mean square of the resulting data points.
Similar to~\cite{ALICE_dielecton13TeV}, additional uncertainties related to the conversion rejection criteria,
the isolation criterion in the ITS and the requirement of a hit in the first ITS layer, as well as on the TPC-ITS matching efficiency, the V0 trigger efficiency and the vertex reconstruction efficiency are added in quadrature. The resulting total systematic uncertainties are 12\% for 
$\mee<0.04$ GeV/$c^2$ and $11\%$ for larger invariant masses, independent of \pTee.
The global 5\% uncertainty resulting from the luminosity measurement is not included in the systematic uncertainties of the data points.

The dielectron measurement is compared with the sum of expected contributions 
from light ($\pi^0$, $\eta$, $\eta$', $\omega$, $\rho$, $\phi$) and heavy-flavor hadron decays within the kinematic range under study. 
The hadronic cocktail is constructed as described in~\cite{ALICE_dielecton13TeV}, with the following exceptions.
The $\pt$ spectrum of $\pi^{\pm}$ in pp collisions at $\sqrt{s}=13$~TeV~\cite{ALICE_PionMult2} is parametrized using a modified Hagedorn function~\cite{HagedornFunc2}. The difference between $\pi^0$ and $\pi^{\pm}$ due to isospin-violating decays, mainly of the $\eta$ meson, is estimated using 
an effective model that describes measured hadron spectra at low \pt and includes strong and electromagnetic decays~\cite{KlausBW}. 
This leads to a \pt-dependent scaling factor applied to the $\pi^{\pm}$ parametrization, which implies an upward shift by 18$\pm$6\% for $\pt\rightarrow0$ that
drops monotonically to below 1\% at $\pt>1$~GeV/$c$. 
The uncertainty of this correction is estimated from variations of the model parameters and propagated into the final cocktail uncertainty.

\begin{figure}[tb]
    \begin{center}
    \includegraphics[width = 0.72\textwidth]{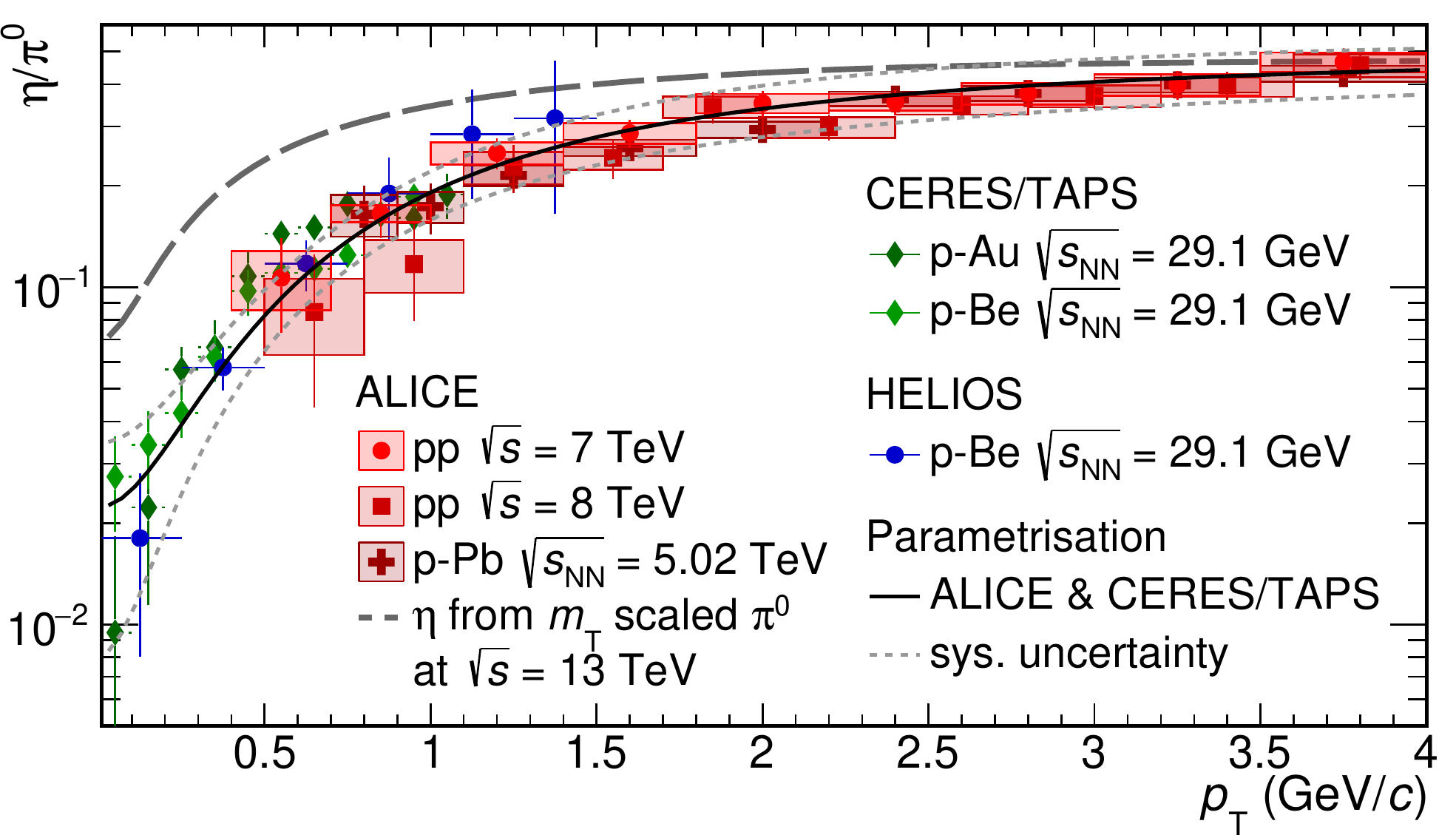}
    \end{center}
    \caption{The ratio $\eta/\pi^0$ as a function of \pt,
    measured in pp and p--A collisions at different center-of-mass energies~\cite{CERES_eta,ALICE_etaPP,ALICE_etaPP8,ALICE_etaPA}. Also shown is the parametrization used for the construction of the hadronic cocktail (solid line), its uncertainty (dotted line), and the expectation from $m_{\rm{T}}$-scaling (dashed line).}
    \label{fig:EtaOverPion}
\end{figure}

The dominant contribution to the hadronic cocktail 
in the kinematic region of interest is given by the $\eta$ meson.
Therefore, a parametrization of the \mbox{ALICE} measurement of $\eta/\pi^0$ ratio as a function of \pt
in pp collisions at $\sqrt{s}=7$~TeV~\cite{ALICE_etaPP}, 8~TeV~\cite{ALICE_etaPP8}, and in p--Pb collisions at $\sqrt{s_{\mathrm {NN}}}=5.02$~TeV~\cite{ALICE_etaPA} is performed and extended to low \pt, using data from CERES/TAPS~\cite{CERES_eta} below $\pt=0.4$~GeV/$c$ and assuming energy independence of the ratio. 
The estimated uncertainty is about 15\% at $\pt>0.5$~GeV/$c$, where data from LHC exist. 
At smaller \pt, a conservative \pt-dependent uncertainty of up to 40\% is assigned, covering the full spread of the data points and a possible weak energy dependence of the $\eta/\pi^0$ ratio. 
The resulting $\eta/\pi^0$ parametrization including the estimated uncertainties is shown in Fig.~\ref{fig:EtaOverPion}. 
It also illustrates that $m_{\rm{ T}}$-scaling~\cite{Hagedorn} fails to describe the measured $\eta/\pi^0$ ratio at low \pt, as reported earlier~\cite{ALICE_etaPP8,mTscaling2}.

The contribution from correlated semileptonic decays of open charm and beauty hadrons is estimated based on the decay distributions from the Perugia 2011 tune of PYTHIA 6.4, normalized to the measured cross sections at midrapidity,
${\rm d}\sigma_{\rm c\overline{c}}/{\rm d}y|_{y=0}=974\pm138\,({\rm stat.})\pm140\,({\rm syst.})\;\mu{\rm b}$ and ${\rm d}\sigma_{\rm b\overline{b}}/{\rm d}y|_{y=0}=79\pm14\,({\rm stat.})\pm 11\,({\rm syst.})\;\mu{\rm b}$,
from the dielectron
analysis in pp collisions at $\sqrt{s}=13$~TeV at nominal field~\cite{ALICE_dielecton13TeV}. Finally, the detector resolution in $\pT$, $\eta_{\rm e}$ and 
azimuthal angle $\varphi_{\rm e}$
is extracted as a function of $\pT$ from the same MC simulation and applied to all decay electrons~\cite{Resolution}. 
To construct the cocktail in intervals of ${\rm d} N_{\rm ch}/{\rm d} \eta$, the light-flavor \pt spectra of the MB cocktail are scaled by the ratio of the 
charged-particle \pt spectra measured in multiplicity intervals to all events having at least one charged particle produced in the pseudorapidity interval \etarange{1} (INEL$>$0 events)~\cite{ALICE_PionMult2}.   
The open-charm contribution is weighted according to the measured enhancement of D mesons at $\pt>1$~GeV/$c$ in pp collisions at $\sqrt{s}=7$ TeV~\cite{ALICE_HFscaling}.  
The overall systematic uncertainties of the hadronic cocktail are estimated by adding in quadrature the uncertainties of the following contributions: the input data parametrizations as a function of $\pt$, the $\pi^0/\pi^{\pm}$ correction factor, the uncertainty of the $\eta/\pi^0$, $\omega/\pi^0$~\cite{PYTHIA8_Monash} and $\rho/\pi^0$~\cite{PYTHIA8_Monash} ratios, the scaling parameters used for $\eta'$~\cite{PYTHIA6} and $\phi$~\cite{PhiScaling}, the branching fractions of the different light-flavor decay channels, the measured cross sections, as well as the estimation of ${\rm d} N_{\rm ch}/{\rm d} \eta$. This results in a systematic uncertainty of the hadronic cocktail between 13$\%$ in the $\pi^0$-Dalitz region and up to $24\%$ in the mass region dominated by the $\eta$ meson.

\begin{figure}[tb]
    \begin{center}
    \includegraphics[width = 0.42\textwidth]{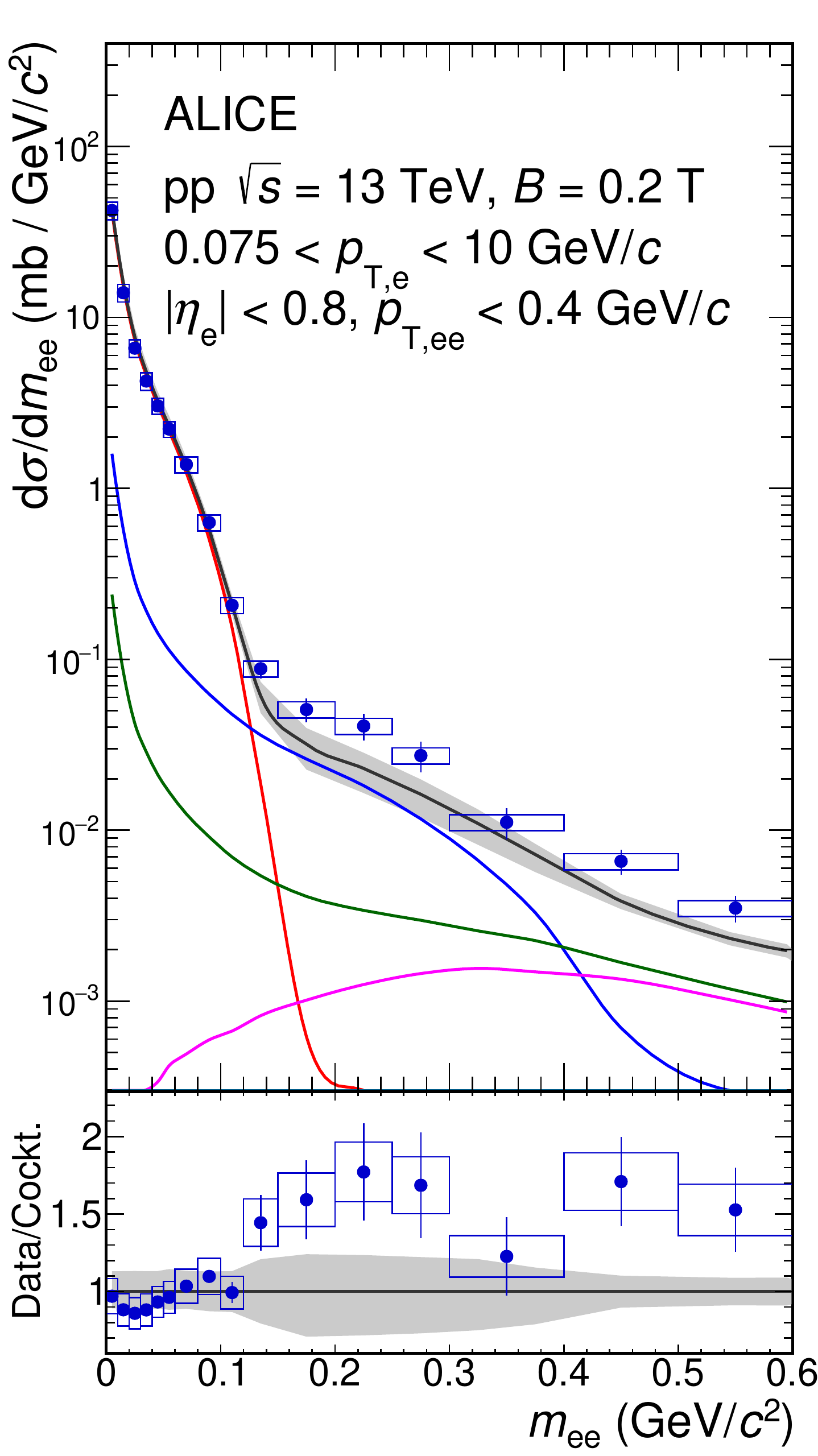}
    \includegraphics[width = 0.42\textwidth]{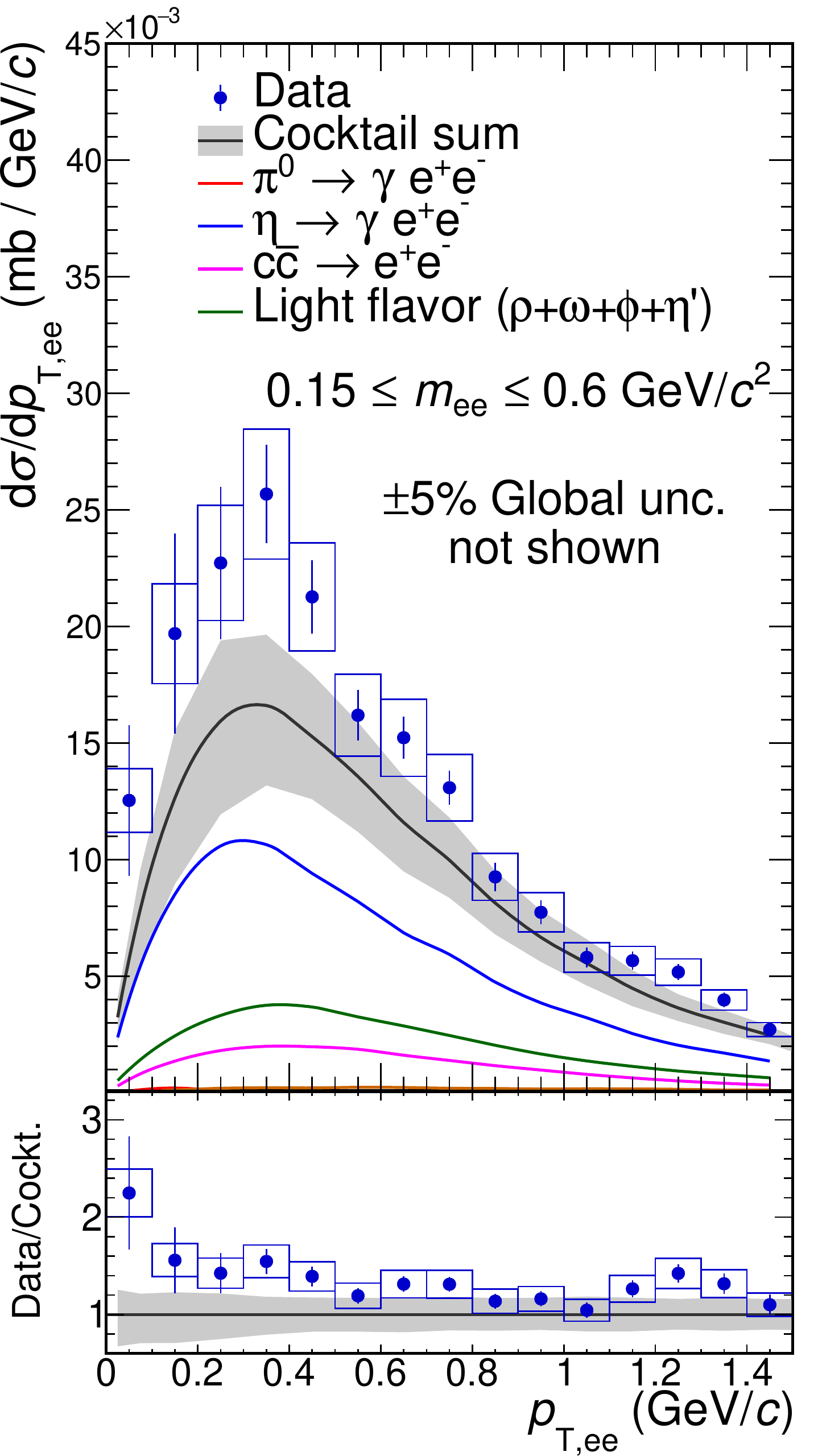}
    \end{center}
    \caption{Differential dielectron cross sections as a function of \mee (left) and \pTee (right). The different components of the hadronic cocktail are shown as solid lines. The error bars and boxes indicate the statistical and systematic uncertainties of the data points. The cocktail uncertainties are shown as gray bands. In the bottom panels, the ratios of data and cocktail are shown.}
    \label{fig:CocktailComp}
\end{figure}

The dielectron cross section as a function of $\mee$ 
in the range $\pTee < 0.4$ GeV/$c$ and within the \mbox{ALICE} single-electron acceptance 
is shown in the left panel of Fig.~\ref{fig:CocktailComp}. 
The data points are compared to the hadronic cocktail. Within the uncertainties, data and cocktail are in good agreement 
at $\mee<m_{\rm \pi}$ while an excess over the hadronic cocktail is observed at larger masses.
The representation of the data as a function of \pTee in the invariant mass region  $0.15<\mee<0.6$ GeV/$c^2$ 
(right panel of Fig.~\ref{fig:CocktailComp})
illustrates that the excess is most pronounced at $\pTee<0.4$ GeV/$c$, while the hadronic cocktail agrees well with the data at higher $\pTee$. In the mass region $0.15<\mee<0.6$~GeV/$c^2$ and for $\pTee < 0.4$ GeV/$c$, the enhancement factor amounts to $1.61\pm 0.13\,(\rm{stat.})\pm 0.17\,(\rm{syst., data}) \pm 0.34\,(\rm{syst., cocktail})$.
The systematic uncertainty is dominated by the uncertainty of the $\eta$ contribution to the hadronic cocktail.

\begin{figure}[tb]
    \begin{center}
    \includegraphics[width = 0.7\textwidth]{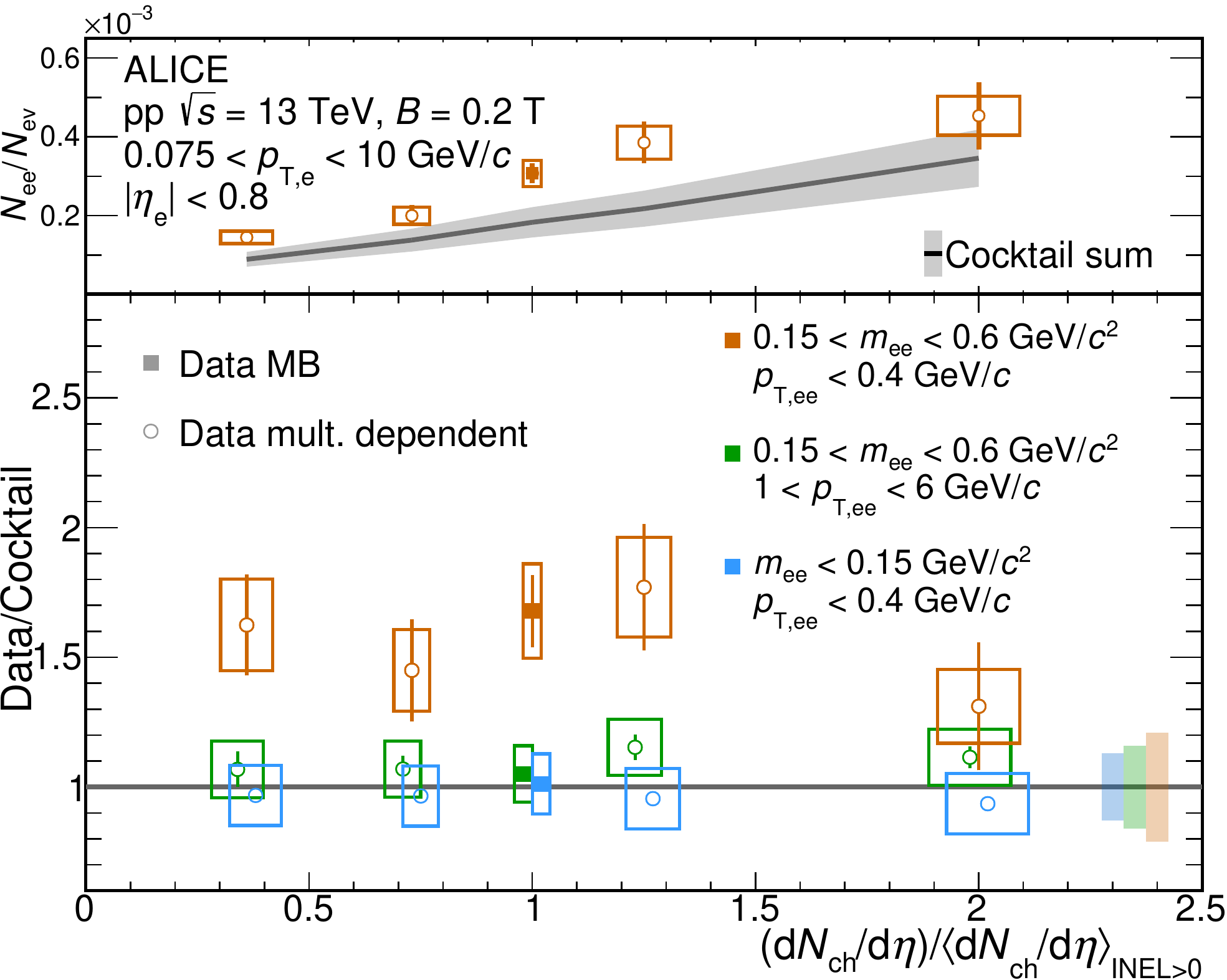}
    \end{center}
    \caption{
    Upper panel: Dielectron yield per event in the excess region as a function of the event multiplicity compared with the hadronic cocktail. Lower panel: Enhancement factor data/cocktail in three different kinematic regions. Error bars and boxes show the statistical and systematic uncertainties of the data points. The cocktail uncertainties are indicated as vertical bars around 1 in the lower panel.
    }
    \label{fig:MultDep}
\end{figure}

The study of the multiplicity dependence of the observed excess may help to unravel the nature of the underlying dielectron production mechanisms~\cite{SoftAnn4}.
To this end, four intervals of the event multiplicity are selected, based on the V0 signal, and the dielectron data are integrated over different regions of $\mee$ and $\pTee$. 
The upper part of Fig.~\ref{fig:MultDep} shows the dielectron yield per event in the interval $0.15 < \mee < 0.6$~GeV/$c^2$ and $\pTee < 0.4$~GeV/$c$ compared with the hadronic cocktail, integrated over the same $\mee$ and $\pTee$ interval, as a function of the relative charged-particle multiplicity at midrapidity, $\left({\rm d} N_{\rm ch}/{\rm d} \eta\right)/\langle {\rm d} N_{\rm ch}/{\rm d} \eta\rangle_{\rm INEL>0}$, where $\langle {\rm d} N_{\rm ch}/{\rm d} \eta\rangle_{\rm INEL>0} = 7.6 \pm 0.5$ is the mean multiplicity in INEL$>$0 pp collisions at $\sqrt{s}=13$~TeV~\cite{ALICE_PionMult}.
The dielectron yield is systematically above the cocktail in all multiplicity intervals. The enhancement of the data over the cocktail 
is shown in the lower part of Fig.~\ref{fig:MultDep}. Within the  experimental accuracy, no clear trend for the multiplicity dependence is found.
Figure ~\ref{fig:MultDep} also shows the multiplicity dependence in control regions at smaller $\mee$ or larger $\pTee$, where no excess is observed.

\begin{figure}[tb]
    \begin{center}
    \includegraphics[width = 0.75\textwidth]{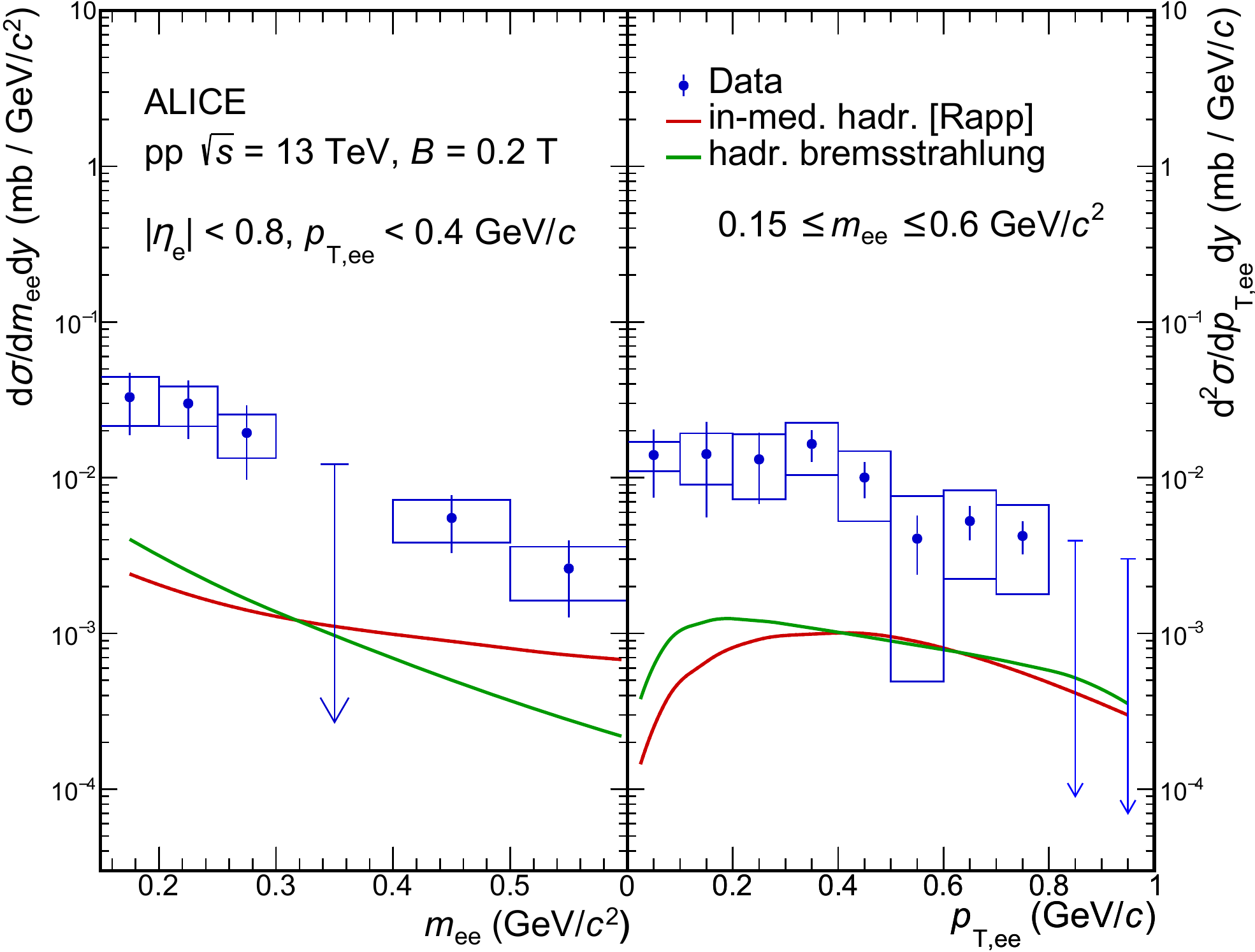}
    \end{center}
    \caption{Dielectron excess spectra 
    as a function of \mee (left) and \pTee(right) after subtraction of the hadronic decay cocktail. The error bars and boxes represent statistical and combined systematic uncertainties from data and cocktail. Arrows indicate upper limits at 90\% confidence level. Also shown as lines are calculations of bremsstrahlung from initial- and final-state hadrons~\cite{Brems_Dielectron}, and thermal dielectron production~\cite{Rapp1,Rapp2,Rapp3}.}
    \label{fig:ExcessSpectra}
\end{figure}

To further characterize the observed dielectron enhancement, the hadronic cocktail is subtracted from the measured $\mee$ and $\pTee$ spectra. 
The extracted excess spectra are corrected for the single-electron acceptance in $\pT$ and $\eta_{\rm e}$, assuming isotropic decay in the pair center-of-mass frame, which enables the measurement of the excess cross section in $\mee > 0.15$~GeV/$c^2$ and $\pTee > 0$ at midrapidity.
The corresponding excess spectra as a function of $\mee$ and $\pTee$ 
are shown in Fig.~\ref{fig:ExcessSpectra}.
The data points are compared with a calculation of bremsstrahlung from initial- and final-state hadrons following the approach in~\cite{Brems_Dielectron} using a mean charge transfer $\langle \Delta Q^2 \rangle =1.32$ in units of the electric charge {\it e} squared and the inelastic hadronic cross section~\cite{ALICE_VanDerMeer}.
Also shown is a calculation of the thermal dielectron yield from a hadronic many-body model~\cite{Rapp1,Rapp2,Rapp3}, assuming a fireball lifetime of 2~fm/$c$, an initual temperature of 216~MeV/$c$ and a freeze-out temperature of 170~MeV/$c$.   
While the hadronic many-body approach is successful in describing the dilepton production in heavy-ion collisions at the SPS~\cite{CeresAuPb2,Na60_InIn}, at RHIC~\cite{Star_AuAu,Star_AuAu2,Phenix_AuAu}, and at the LHC~\cite{ALICE_dielectronPbPb}, it fails to describe the present dielectron results in pp collisions. 
An enhancement of dielectrons at very low \pTee  in peripheral Au-Au collisions at $\sqrt{s_{\mathrm {NN}}}=200$~GeV, reported by the STAR collaboration~\cite{StarViaPhotonProduction}, could be explained by coherent two-photon production of lepton pairs in the strong electric fields of the colliding nuclei~\cite{Spencer,CoherentPhoton,CoherentPhoton2}. Owing to the strong $Z$-dependence, this mechanism is not sufficient to describe the present enhancement in pp collisions. 

The  results reported here are expected to encourage further theoretical work.

In conclusion, an excess of soft dielectrons over the expectation from hadron decays is observed in pp collisions at \mbox{$\sqrt{s}=13$~TeV}. The enhancement factor shows no dependence on the event multiplicity, and the acceptance-corrected excess yield cannot be explained by bremsstrahlung from initial- and final-state hadrons or by thermal dielectron production. The excess of soft dielectrons in pp is an intriguing observation,
although its significance is presently limited to 1.6$\sigma$, mostly by the  uncertainty of the hadronic cocktail. Forthcoming precision measurements with the upgraded ALICE detector will help to further elucidate this finding, including a possible connection to earlier observations of anomalous soft-photon and soft-dielectron production at lower collision energies.


\newenvironment{acknowledgement}{\relax}{\relax}
\begin{acknowledgement}
\section*{Acknowledgements}

The ALICE Collaboration would like to thank all its engineers and technicians for their invaluable contributions to the construction of the experiment and the CERN accelerator teams for the outstanding performance of the LHC complex.
The ALICE Collaboration gratefully acknowledges the resources and support provided by all Grid centres and the Worldwide LHC Computing Grid (WLCG) collaboration.
The ALICE Collaboration acknowledges the following funding agencies for their support in building and running the ALICE detector:
A. I. Alikhanyan National Science Laboratory (Yerevan Physics Institute) Foundation (ANSL), State Committee of Science and World Federation of Scientists (WFS), Armenia;
Austrian Academy of Sciences, Austrian Science Fund (FWF): [M 2467-N36] and Nationalstiftung f\"{u}r Forschung, Technologie und Entwicklung, Austria;
Ministry of Communications and High Technologies, National Nuclear Research Center, Azerbaijan;
Conselho Nacional de Desenvolvimento Cient\'{\i}fico e Tecnol\'{o}gico (CNPq), Financiadora de Estudos e Projetos (Finep), Funda\c{c}\~{a}o de Amparo \`{a} Pesquisa do Estado de S\~{a}o Paulo (FAPESP) and Universidade Federal do Rio Grande do Sul (UFRGS), Brazil;
Ministry of Education of China (MOEC) , Ministry of Science \& Technology of China (MSTC) and National Natural Science Foundation of China (NSFC), China;
Ministry of Science and Education and Croatian Science Foundation, Croatia;
Centro de Aplicaciones Tecnol\'{o}gicas y Desarrollo Nuclear (CEADEN), Cubaenerg\'{\i}a, Cuba;
Ministry of Education, Youth and Sports of the Czech Republic, Czech Republic;
The Danish Council for Independent Research | Natural Sciences, the VILLUM FONDEN and Danish National Research Foundation (DNRF), Denmark;
Helsinki Institute of Physics (HIP), Finland;
Commissariat \`{a} l'Energie Atomique (CEA) and Institut National de Physique Nucl\'{e}aire et de Physique des Particules (IN2P3) and Centre National de la Recherche Scientifique (CNRS), France;
Bundesministerium f\"{u}r Bildung und Forschung (BMBF) and GSI Helmholtzzentrum f\"{u}r Schwerionenforschung GmbH, Germany;
General Secretariat for Research and Technology, Ministry of Education, Research and Religions, Greece;
National Research, Development and Innovation Office, Hungary;
Department of Atomic Energy Government of India (DAE), Department of Science and Technology, Government of India (DST), University Grants Commission, Government of India (UGC) and Council of Scientific and Industrial Research (CSIR), India;
Indonesian Institute of Science, Indonesia;
Centro Fermi - Museo Storico della Fisica e Centro Studi e Ricerche Enrico Fermi and Istituto Nazionale di Fisica Nucleare (INFN), Italy;
Institute for Innovative Science and Technology , Nagasaki Institute of Applied Science (IIST), Japanese Ministry of Education, Culture, Sports, Science and Technology (MEXT) and Japan Society for the Promotion of Science (JSPS) KAKENHI, Japan;
Consejo Nacional de Ciencia (CONACYT) y Tecnolog\'{i}a, through Fondo de Cooperaci\'{o}n Internacional en Ciencia y Tecnolog\'{i}a (FONCICYT) and Direcci\'{o}n General de Asuntos del Personal Academico (DGAPA), Mexico;
Nederlandse Organisatie voor Wetenschappelijk Onderzoek (NWO), Netherlands;
The Research Council of Norway, Norway;
Commission on Science and Technology for Sustainable Development in the South (COMSATS), Pakistan;
Pontificia Universidad Cat\'{o}lica del Per\'{u}, Peru;
Ministry of Science and Higher Education, National Science Centre and WUT ID-UB, Poland;
Korea Institute of Science and Technology Information and National Research Foundation of Korea (NRF), Republic of Korea;
Ministry of Education and Scientific Research, Institute of Atomic Physics and Ministry of Research and Innovation and Institute of Atomic Physics, Romania;
Joint Institute for Nuclear Research (JINR), Ministry of Education and Science of the Russian Federation, National Research Centre Kurchatov Institute, Russian Science Foundation and Russian Foundation for Basic Research, Russia;
Ministry of Education, Science, Research and Sport of the Slovak Republic, Slovakia;
National Research Foundation of South Africa, South Africa;
Swedish Research Council (VR) and Knut \& Alice Wallenberg Foundation (KAW), Sweden;
European Organization for Nuclear Research, Switzerland;
Suranaree University of Technology (SUT), National Science and Technology Development Agency (NSDTA) and Office of the Higher Education Commission under NRU project of Thailand, Thailand;
Turkish Atomic Energy Agency (TAEK), Turkey;
National Academy of  Sciences of Ukraine, Ukraine;
Science and Technology Facilities Council (STFC), United Kingdom;
National Science Foundation of the United States of America (NSF) and United States Department of Energy, Office of Nuclear Physics (DOE NP), United States of America.
\end{acknowledgement}

\bibliographystyle{utphys}   
\bibliography{bibliography}

\newpage
\appendix

%
%

\section{The ALICE Collaboration}
\label{app:collab}

\begingroup
\small
\begin{flushleft}
S.~Acharya\Irefn{org141}\And 
D.~Adamov\'{a}\Irefn{org95}\And 
A.~Adler\Irefn{org74}\And 
J.~Adolfsson\Irefn{org81}\And 
M.M.~Aggarwal\Irefn{org100}\And 
G.~Aglieri Rinella\Irefn{org34}\And 
M.~Agnello\Irefn{org30}\And 
N.~Agrawal\Irefn{org10}\textsuperscript{,}\Irefn{org54}\And 
Z.~Ahammed\Irefn{org141}\And 
S.~Ahmad\Irefn{org16}
% ...
\And
\renewcommand\labelenumi{\textsuperscript{\theenumi}~}

\section*{Affiliation notes}
\renewcommand\theenumi{\roman{enumi}}
\begin{Authlist}
\item \Adef{org*}Deceased
\item \Adef{orgI}Italian National Agency for New Technologies, Energy and Sustainable Economic Development (ENEA), Bologna, Italy
\item \Adef{orgII}Dipartimento DET del Politecnico di Torino, Turin, Italy
\item \Adef{orgIII}M.V. Lomonosov Moscow State University, D.V. Skobeltsyn Institute of Nuclear, Physics, Moscow, Russia
\item \Adef{orgIV}Department of Applied Physics, Aligarh Muslim University, Aligarh, India
\item \Adef{orgV}Institute of Theoretical Physics, University of Wroclaw, Poland
\end{Authlist}

\section*{Collaboration Institutes}
\renewcommand\theenumi{\arabic{enumi}~}
\begin{Authlist}
\item \Idef{org1}A.I. Alikhanyan National Science Laboratory (Yerevan Physics Institute) Foundation, Yerevan, Armenia
\item \Idef{org2}Bogolyubov Institute for Theoretical Physics, National Academy of Sciences of Ukraine, Kiev, Ukraine
\item \Idef{org3}Bose Institute, Department of Physics  and Centre for Astroparticle Physics and Space Science (CAPSS), Kolkata, India
\item \Idef{org4}Budker Institute for Nuclear Physics, Novosibirsk, Russia
\item \Idef{org5}California Polytechnic State University, San Luis Obispo, California, United States
\item \Idef{org6}Central China Normal University, Wuhan, China
\item \Idef{org7}Centre de Calcul de l'IN2P3, Villeurbanne, Lyon, France
\item \Idef{org8}Centro de Aplicaciones Tecnol\'{o}gicas y Desarrollo Nuclear (CEADEN), Havana, Cuba
\item \Idef{org9}Centro de Investigaci\'{o}n y de Estudios Avanzados (CINVESTAV), Mexico City and M\'{e}rida, Mexico
\item \Idef{org10}Centro Fermi - Museo Storico della Fisica e Centro Studi e Ricerche ``Enrico Fermi', Rome, Italy
\item \Idef{org11}Chicago State University, Chicago, Illinois, United States
\item \Idef{org12}China Institute of Atomic Energy, Beijing, China
\item \Idef{org13}Comenius University Bratislava, Faculty of Mathematics, Physics and Informatics, Bratislava, Slovakia
\item \Idef{org14}COMSATS University Islamabad, Islamabad, Pakistan
\item \Idef{org15}Creighton University, Omaha, Nebraska, United States
\item \Idef{org16}Department of Physics, Aligarh Muslim University, Aligarh, India
\item \Idef{org17}Department of Physics, Pusan National University, Pusan, Republic of Korea
\item \Idef{org18}Department of Physics, Sejong University, Seoul, Republic of Korea
\item \Idef{org19}Department of Physics, University of California, Berkeley, California, United States
\item \Idef{org20}Department of Physics, University of Oslo, Oslo, Norway
\item \Idef{org21}Department of Physics and Technology, University of Bergen, Bergen, Norway
\item \Idef{org22}Dipartimento di Fisica dell'Universit\`{a} 'La Sapienza' and Sezione INFN, Rome, Italy
\item \Idef{org23}Dipartimento di Fisica dell'Universit\`{a} and Sezione INFN, Cagliari, Italy
\item \Idef{org24}Dipartimento di Fisica dell'Universit\`{a} and Sezione INFN, Trieste, Italy
\item \Idef{org25}Dipartimento di Fisica dell'Universit\`{a} and Sezione INFN, Turin, Italy
\item \Idef{org26}Dipartimento di Fisica e Astronomia dell'Universit\`{a} and Sezione INFN, Bologna, Italy
\item \Idef{org27}Dipartimento di Fisica e Astronomia dell'Universit\`{a} and Sezione INFN, Catania, Italy
\item \Idef{org28}Dipartimento di Fisica e Astronomia dell'Universit\`{a} and Sezione INFN, Padova, Italy
\item \Idef{org29}Dipartimento di Fisica `E.R.~Caianiello' dell'Universit\`{a} and Gruppo Collegato INFN, Salerno, Italy
\item \Idef{org30}Dipartimento DISAT del Politecnico and Sezione INFN, Turin, Italy
\item \Idef{org31}Dipartimento di Scienze e Innovazione Tecnologica dell'Universit\`{a} del Piemonte Orientale and INFN Sezione di Torino, Alessandria, Italy
\item \Idef{org32}Dipartimento di Scienze MIFT, Universit\`{a} di Messina, Messina, Italy
\item \Idef{org33}Dipartimento Interateneo di Fisica `M.~Merlin' and Sezione INFN, Bari, Italy
\item \Idef{org34}European Organization for Nuclear Research (CERN), Geneva, Switzerland
\item \Idef{org35}Faculty of Electrical Engineering, Mechanical Engineering and Naval Architecture, University of Split, Split, Croatia
\item \Idef{org36}Faculty of Engineering and Science, Western Norway University of Applied Sciences, Bergen, Norway
\item \Idef{org37}Faculty of Nuclear Sciences and Physical Engineering, Czech Technical University in Prague, Prague, Czech Republic
\item \Idef{org38}Faculty of Science, P.J.~\v{S}af\'{a}rik University, Ko\v{s}ice, Slovakia
\item \Idef{org39}Frankfurt Institute for Advanced Studies, Johann Wolfgang Goethe-Universit\"{a}t Frankfurt, Frankfurt, Germany
\item \Idef{org40}Fudan University, Shanghai, China
\item \Idef{org41}Gangneung-Wonju National University, Gangneung, Republic of Korea
\item \Idef{org42}Gauhati University, Department of Physics, Guwahati, India
\item \Idef{org43}Helmholtz-Institut f\"{u}r Strahlen- und Kernphysik, Rheinische Friedrich-Wilhelms-Universit\"{a}t Bonn, Bonn, Germany
\item \Idef{org44}Helsinki Institute of Physics (HIP), Helsinki, Finland
\item \Idef{org45}High Energy Physics Group,  Universidad Aut\'{o}noma de Puebla, Puebla, Mexico
\item \Idef{org46}Hiroshima University, Hiroshima, Japan
\item \Idef{org47}Hochschule Worms, Zentrum  f\"{u}r Technologietransfer und Telekommunikation (ZTT), Worms, Germany
\item \Idef{org48}Horia Hulubei National Institute of Physics and Nuclear Engineering, Bucharest, Romania
\item \Idef{org49}Indian Institute of Technology Bombay (IIT), Mumbai, India
\item \Idef{org50}Indian Institute of Technology Indore, Indore, India
\item \Idef{org51}Indonesian Institute of Sciences, Jakarta, Indonesia
\item \Idef{org52}INFN, Laboratori Nazionali di Frascati, Frascati, Italy
\item \Idef{org53}INFN, Sezione di Bari, Bari, Italy
\item \Idef{org54}INFN, Sezione di Bologna, Bologna, Italy
\item \Idef{org55}INFN, Sezione di Cagliari, Cagliari, Italy
\item \Idef{org56}INFN, Sezione di Catania, Catania, Italy
\item \Idef{org57}INFN, Sezione di Padova, Padova, Italy
\item \Idef{org58}INFN, Sezione di Roma, Rome, Italy
\item \Idef{org59}INFN, Sezione di Torino, Turin, Italy
\item \Idef{org60}INFN, Sezione di Trieste, Trieste, Italy
\item \Idef{org61}Inha University, Incheon, Republic of Korea
\item \Idef{org62}Institute for Nuclear Research, Academy of Sciences, Moscow, Russia
\item \Idef{org63}Institute for Subatomic Physics, Utrecht University/Nikhef, Utrecht, Netherlands
\item \Idef{org64}Institute of Experimental Physics, Slovak Academy of Sciences, Ko\v{s}ice, Slovakia
\item \Idef{org65}Institute of Physics, Homi Bhabha National Institute, Bhubaneswar, India
\item \Idef{org66}Institute of Physics of the Czech Academy of Sciences, Prague, Czech Republic
\item \Idef{org67}Institute of Space Science (ISS), Bucharest, Romania
\item \Idef{org68}Institut f\"{u}r Kernphysik, Johann Wolfgang Goethe-Universit\"{a}t Frankfurt, Frankfurt, Germany
\item \Idef{org69}Instituto de Ciencias Nucleares, Universidad Nacional Aut\'{o}noma de M\'{e}xico, Mexico City, Mexico
\item \Idef{org70}Instituto de F\'{i}sica, Universidade Federal do Rio Grande do Sul (UFRGS), Porto Alegre, Brazil
\item \Idef{org71}Instituto de F\'{\i}sica, Universidad Nacional Aut\'{o}noma de M\'{e}xico, Mexico City, Mexico
\item \Idef{org72}iThemba LABS, National Research Foundation, Somerset West, South Africa
\item \Idef{org73}Jeonbuk National University, Jeonju, Republic of Korea
\item \Idef{org74}Johann-Wolfgang-Goethe Universit\"{a}t Frankfurt Institut f\"{u}r Informatik, Fachbereich Informatik und Mathematik, Frankfurt, Germany
\item \Idef{org75}Joint Institute for Nuclear Research (JINR), Dubna, Russia
\item \Idef{org76}Korea Institute of Science and Technology Information, Daejeon, Republic of Korea
\item \Idef{org77}KTO Karatay University, Konya, Turkey
\item \Idef{org78}Laboratoire de Physique des 2 Infinis, Ir\`{e}ne Joliot-Curie, Orsay, France
\item \Idef{org79}Laboratoire de Physique Subatomique et de Cosmologie, Universit\'{e} Grenoble-Alpes, CNRS-IN2P3, Grenoble, France
\item \Idef{org80}Lawrence Berkeley National Laboratory, Berkeley, California, United States
\item \Idef{org81}Lund University Department of Physics, Division of Particle Physics, Lund, Sweden
\item \Idef{org82}Nagasaki Institute of Applied Science, Nagasaki, Japan
\item \Idef{org83}Nara Women{'}s University (NWU), Nara, Japan
\item \Idef{org84}National and Kapodistrian University of Athens, School of Science, Department of Physics , Athens, Greece
\item \Idef{org85}National Centre for Nuclear Research, Warsaw, Poland
\item \Idef{org86}National Institute of Science Education and Research, Homi Bhabha National Institute, Jatni, India
\item \Idef{org87}National Nuclear Research Center, Baku, Azerbaijan
\item \Idef{org88}National Research Centre Kurchatov Institute, Moscow, Russia
\item \Idef{org89}Niels Bohr Institute, University of Copenhagen, Copenhagen, Denmark
\item \Idef{org90}Nikhef, National institute for subatomic physics, Amsterdam, Netherlands
\item \Idef{org91}NRC Kurchatov Institute IHEP, Protvino, Russia
\item \Idef{org92}NRC \guillemotleft Kurchatov\guillemotright~Institute - ITEP, Moscow, Russia
\item \Idef{org93}NRNU Moscow Engineering Physics Institute, Moscow, Russia
\item \Idef{org94}Nuclear Physics Group, STFC Daresbury Laboratory, Daresbury, United Kingdom
\item \Idef{org95}Nuclear Physics Institute of the Czech Academy of Sciences, \v{R}e\v{z} u Prahy, Czech Republic
\item \Idef{org96}Oak Ridge National Laboratory, Oak Ridge, Tennessee, United States
\item \Idef{org97}Ohio State University, Columbus, Ohio, United States
\item \Idef{org98}Petersburg Nuclear Physics Institute, Gatchina, Russia
\item \Idef{org99}Physics department, Faculty of science, University of Zagreb, Zagreb, Croatia
\item \Idef{org100}Physics Department, Panjab University, Chandigarh, India
\item \Idef{org101}Physics Department, University of Jammu, Jammu, India
\item \Idef{org102}Physics Department, University of Rajasthan, Jaipur, India
\item \Idef{org103}Physikalisches Institut, Eberhard-Karls-Universit\"{a}t T\"{u}bingen, T\"{u}bingen, Germany
\item \Idef{org104}Physikalisches Institut, Ruprecht-Karls-Universit\"{a}t Heidelberg, Heidelberg, Germany
\item \Idef{org105}Physik Department, Technische Universit\"{a}t M\"{u}nchen, Munich, Germany
\item \Idef{org106}Politecnico di Bari, Bari, Italy
\item \Idef{org107}Research Division and ExtreMe Matter Institute EMMI, GSI Helmholtzzentrum f\"ur Schwerionenforschung GmbH, Darmstadt, Germany
\item \Idef{org108}Rudjer Bo\v{s}kovi\'{c} Institute, Zagreb, Croatia
\item \Idef{org109}Russian Federal Nuclear Center (VNIIEF), Sarov, Russia
\item \Idef{org110}Saha Institute of Nuclear Physics, Homi Bhabha National Institute, Kolkata, India
\item \Idef{org111}School of Physics and Astronomy, University of Birmingham, Birmingham, United Kingdom
\item \Idef{org112}Secci\'{o}n F\'{\i}sica, Departamento de Ciencias, Pontificia Universidad Cat\'{o}lica del Per\'{u}, Lima, Peru
\item \Idef{org113}St. Petersburg State University, St. Petersburg, Russia
\item \Idef{org114}Stefan Meyer Institut f\"{u}r Subatomare Physik (SMI), Vienna, Austria
\item \Idef{org115}SUBATECH, IMT Atlantique, Universit\'{e} de Nantes, CNRS-IN2P3, Nantes, France
\item \Idef{org116}Suranaree University of Technology, Nakhon Ratchasima, Thailand
\item \Idef{org117}Technical University of Ko\v{s}ice, Ko\v{s}ice, Slovakia
\item \Idef{org118}The Henryk Niewodniczanski Institute of Nuclear Physics, Polish Academy of Sciences, Cracow, Poland
\item \Idef{org119}The University of Texas at Austin, Austin, Texas, United States
\item \Idef{org120}Universidad Aut\'{o}noma de Sinaloa, Culiac\'{a}n, Mexico
\item \Idef{org121}Universidade de S\~{a}o Paulo (USP), S\~{a}o Paulo, Brazil
\item \Idef{org122}Universidade Estadual de Campinas (UNICAMP), Campinas, Brazil
\item \Idef{org123}Universidade Federal do ABC, Santo Andre, Brazil
\item \Idef{org124}University of Cape Town, Cape Town, South Africa
\item \Idef{org125}University of Houston, Houston, Texas, United States
\item \Idef{org126}University of Jyv\"{a}skyl\"{a}, Jyv\"{a}skyl\"{a}, Finland
\item \Idef{org127}University of Liverpool, Liverpool, United Kingdom
\item \Idef{org128}University of Science and Technology of China, Hefei, China
\item \Idef{org129}University of South-Eastern Norway, Tonsberg, Norway
\item \Idef{org130}University of Tennessee, Knoxville, Tennessee, United States
\item \Idef{org131}University of the Witwatersrand, Johannesburg, South Africa
\item \Idef{org132}University of Tokyo, Tokyo, Japan
\item \Idef{org133}University of Tsukuba, Tsukuba, Japan
\item \Idef{org134}Universit\'{e} Clermont Auvergne, CNRS/IN2P3, LPC, Clermont-Ferrand, France
\item \Idef{org135}Universit\'{e} de Lyon, Universit\'{e} Lyon 1, CNRS/IN2P3, IPN-Lyon, Villeurbanne, Lyon, France
\item \Idef{org136}Universit\'{e} de Strasbourg, CNRS, IPHC UMR 7178, F-67000 Strasbourg, France, Strasbourg, France
\item \Idef{org137}Universit\'{e} Paris-Saclay Centre d'Etudes de Saclay (CEA), IRFU, D\'{e}partment de Physique Nucl\'{e}aire (DPhN), Saclay, France
\item \Idef{org138}Universit\`{a} degli Studi di Foggia, Foggia, Italy
\item \Idef{org139}Universit\`{a} degli Studi di Pavia, Pavia, Italy
\item \Idef{org140}Universit\`{a} di Brescia, Brescia, Italy
\item \Idef{org141}Variable Energy Cyclotron Centre, Homi Bhabha National Institute, Kolkata, India
\item \Idef{org142}Warsaw University of Technology, Warsaw, Poland
\item \Idef{org143}Wayne State University, Detroit, Michigan, United States
\item \Idef{org144}Westf\"{a}lische Wilhelms-Universit\"{a}t M\"{u}nster, Institut f\"{u}r Kernphysik, M\"{u}nster, Germany
\item \Idef{org145}Wigner Research Centre for Physics, Budapest, Hungary
\item \Idef{org146}Yale University, New Haven, Connecticut, United States
\item \Idef{org147}Yonsei University, Seoul, Republic of Korea
\end{Authlist}
\endgroup
  
\end{document}